# Semi-supervised GAN for Classification of Multispectral Imagery Acquired by UAVs

Hamideh Kerdegari[1], Manzoor Razaak[1], Vasileios Argyriou[1], Paolo Remagnino[1]

*Abstract*— Unmanned aerial vehicles (UAV) are used in precision agriculture (PA) to enable aerial monitoring of farmlands. Intelligent methods are required to pinpoint weed infestations and make optimal choice of pesticide. UAV can fly a multispectral camera and collect data. However, the classification of multispectral images using supervised machine learning algorithms such as convolutional neural networks (CNN) requires large amount of training data. This is a common drawback in deep learning we try to circumvent making use of a semi-supervised generative adversarial networks (GAN), providing a pixel-wise classification for all the acquired multispectral images. Our algorithm consists of a generator network that provides photo-realistic images as extra training data to a multi-class classifier, acting as a discriminator and trained on small amounts of labeled data. The performance of the proposed method is evaluated on the weedNet dataset consisting of multispectral crop and weed images collected by a micro aerial vehicle (MAV). The results by the proposed semi-supervised GAN achieves high classification accuracy and demonstrates the potential of GAN-based methods for the challenging task of multispectral image classification.

## I. INTRODUCTION

Weed infestation is a major challenge for the agriculture sector. Traditional methods of weed removal are time consuming: they require farmers to physically survey, identify and treat the infested areas. Recently, farmers have used UAV with cameras (RGB, multi or hyper spectral) to obtain a better view of their farm and identify specific weed infestations.
RGB cameras and multispectral sensors flown by UAV have proven to be useful in early weed detection. Studies that utilize RGB cameras mainly apply feature extraction techniques for the detection and classification of weed from crop [1], [2]. Although RGB image analysis methods can be successfully used for weed identification and classification [3], RGB images captured by UAV have few limitations in crop-weed disambiguation such as capturing less information at higher altitude and decreasing vegetation indices as altitude increases [4]. Due to the limitations of the visible spectrum (RGB band), multispectral cameras with additional spectral bands have shown to provide detailed information that enables accurate calculation of vegetation indices and crop-weed classification [5]. Studies employing multispectral cameras apply different approaches for weed detection such as Mahalanobis distance computation between vegetation rates [6], partial least squares discriminant analysis classification models [7], support vector machine (SVM) [8] and decision tree (DT) methods [9], where multispectral images were classified along with the use of normalized difference vegetation indices (NDVI) thresholding.

More recently, deep learning methods have been explored for crop-weed disambiguation. For instance, Sa et al. [10] applied cascaded CNN, Segnet, on multispectral image datasets for classifying sugar beet crop from weed. They trained six models on different spectral channels and achieved a classification F1 score of 0.8. The study was further extended to include a sliding window approach on orthomosaic maps of the farm to apply a deep neural network and achieved improved performance accuracy [11]. Similarly, several other studies applied CNN based methods for weed classification with images captured from both UAV and ground based vehicles and achieved a high performance accuracy [12], [13]. However, a large amount of training data for learning is an inherent requirement of deep learning methods. The lack of large corpora of specific labeled data is a challenge in general and for multispectral data in particular. Furthermore, collecting large corpora of multispectral image data with UAV platforms for crop-weed classification system is time consuming and expensive. To address this challenge, this paper utilizes a semi-supervised version of the generative adversarial networks (GAN) [14]. In the presented GAN based semi-supervised classification method, a generator creates large realistic images, in turn, forcing a discriminator to learn better features for more accurate pixel classification. To the best of our knowledge, application of GAN methods for multispectral image classification is not well explored and our work addresses this research gap. Next section describes the semi-supervised GAN and its structure, and then results and conclusion are presented in Section III and IV, respectively.

## II. SEMI-SUPERVISED GENERATIVE ADVERSARIAL NETWORKS

The GAN framework was first introduced by Goodfellow et al. [14] to train deep generative models. A GAN usually contains two networks: a generative *(G)* network and a discriminative *(D)* network. Both networks *G* and *D* are trained simultaneously in an adversarial manner, where *G* tries to generate fake inputs as real as possible, and *D* tries to disambiguate between real and fake data. The following

This work is part of the 5GRIT project supported by the Department for Digital, Culture, Media and Sport (DCMS), UK, through their 5G Testbeds Program.

[1]The Robot Vision Team (RoViT), Department of Computer Science, Kingston University, Kingston Upon Thames, UK. Corresponding author's: h.kerdegari@kingston.ac.uk

formulation shows *G* and *D* competition in a two-player minmax game with value function *V (D,G)*:

$$\min_G \max_D V(D,G) = E_{x \sim p(x)}[\log D(x)] + E_{z \sim p(z)}[\log(1 - D(G(z)))] \quad (1)$$

Where symbol *E* represents the expected value. *G* transforms a noise variable *z* into *G(z)*, that is a sample from distribution $p_z$, and distribution $p_z$ should converge to distribution $p_x$. *D* is trained to minimize *logD(x)* while *G* is trained to minimize *log(1 - D(G(z)))*.

Unlike the canonical GAN, where the discriminator is a binary classifier for discriminating real and fake images, semi-supervised GAN implements a multiclass classifier. This paper extends the canonical GAN by replacing the traditional discriminator *D* with a fully convolutional multiclass classifier, which, instead, of predicting whether a sample *x* belongs to the data distribution (real or fake), it assigns to each input image pixel *a* label *y* from the *n* classes (i.e. crop, weed or background) or mark it as a fake sample (extra *n + 1* class). More specifically, *D* network predicts the confidence for *n* classes of image pixels and softmax is employed to obtain the probability of sample *x* belonging to each class. Figure 1 presents a schematic description of the semi-supervised GAN architecture. Note that our GAN formulation is different from the typical GAN, where the discriminator is a binary classifier for discriminating real/fake images, while our discriminator performs multiclass pixel categorization.

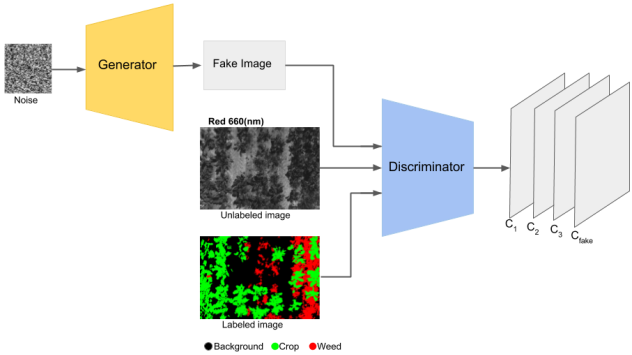

Fig. 1. The semi-supervised GAN architecture. Random noise is used by the Generator to create an image. The Discriminator uses generated data, unlabeled and labeled data to learn class confidence and produces confidence maps for each class as well as a label for fake data.

*A. System overview*

Here, we present a summary of our semi-supervised GAN architecture including both generator and discriminator. The generator network, shown in Figure 2, takes a uniform noise distribution as input, followed by a series of four convolution layers and generates a fake image resembling samples from real data distribution. The discriminator network processes the generated images, unlabeled images and a small number of labeled multi-spectral images to learn class confidence, producing a confidence map for each class as well as a label for fake data. The underlying idea is that adding large fake multispectral images forces real samples to be close in the feature space, which, in turn, improves classification accuracy. Note that the proposed semi-supervised GAN adopts the DCGAN [15] architecture with a modification in the last layer of the discriminator, replacing the sigmoid activation function with the softmax to enable pixel-wise multiclass classification. All the networks are implemented using the Keras library with a Tensorflow backend. The standard Adam optimizer with momentum is used for the discriminator and the generator optimization with learning rate and momentum ($\_1$) set to 0.0002 and 0.5, respectively. A batch size of 32 and batch normalization are utilized for both networks. The ReLU activation function is applied in the generator for all layers except for the output, which uses the Tanh and the LeakyReLU activation in the discriminator for all layers. In the experiments, no data augmentation or post-processing is performed.

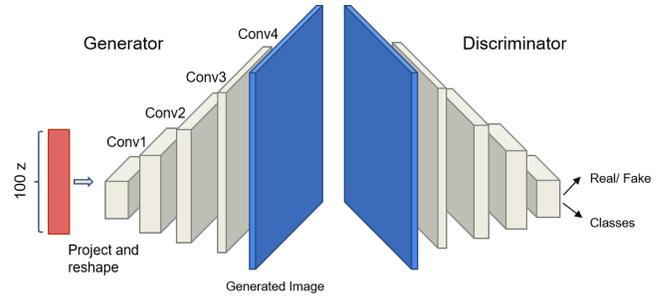

Fig. 2. The network architecture of our semi-supervised GAN. The noise is a vector of size 100 sampled from a uniform distribution and is used as input to the generator. The number of feature maps in the four different convolutional layers, respectively, are 256, 128, 64, 32 and 1.

III. EXPERIMENTAL RESULTS

The proposed method is evaluated on the weedNet [10] dataset collected by a micro aerial vehicle (MAV) equipped with a 4-band Sequoia multispectral camera. The multispectral images are captured from sugar beets field at 2 meter height. The dataset contains only NIR and Red channel due to difficulties in image registration of other bands. From corresponding NIR and Red channel images, the Normalized NDVI is extracted indicating the difference between soil and plant. Therefore, each training/test image consists of the 790nm NIR channel, the 660nm Red channel, and NDVI imagery. For semi-supervised training, different percentages of pixel-wise annotated images (such as 50%, 40% and 30%) are used as labeled data to the discriminator and the rest of images are without pixel-wise annotations.

Quantitative results of our method on weedNet are shown in Table I. F1 measure with a varying number of input channels and different amount of labeled data are used as evaluation metric. Considering the difficulty of the dataset, all models (including different channels + different amount of labeled data) perform reasonably well (about 80% for all classes). As shown in Table I, two input channels (Red and NIR) yield higher performance compared to single channels as they contain more useful features to be used by the semi-supervised

GAN network. However, using 3 channels (NDVI+Red+NIR) did not improve performance as NDVI depends on NIR and Red channels rather than capturing new information. Furthermore, the network was evaluated by reducing the amount of labeled data starting at 50% and then reducing by step 10 to 30% to find out how it affects the classification performance. It is expected that higher amount of labeled data result in better performance. It can be seen by comparing the results of the 50%, 40% and 30% in Table I.

TABLE I
Results on the weedNet dataset using 50%, 40% and 30% of labeled data with different number of channels for semi-supervised GAN. Higher F1 values indicate better classification performance.

| F1 Score | Semi-supervised GAN | | | | | |
|---|---|---|---|---|---|---|
| Labeled Data | 50% | | 40% | | 30% | |
| Channel | Crop | Weed | Crop | Weed | Crop | Weed |
| Red | 0.831 | 0.814 | 0.822 | 0.813 | 0.792 | 0.813 |
| NIR | 0.839 | 0.823 | 0.80 | 0.821 | 0.782 | 0.733 |
| NDVI | 0.826 | 0.803 | 0.817 | 0.79 | 0.788 | 0.812 |
| Red+NIR | **0.857** | **0.865** | 0.837 | 0.834 | 0.823 | 0.815 |
| Red+NIR+NDVI | 0.852 | 0.831 | 0.847 | 0.821 | 0.816 | 0.812 |

IV. CONCLUSION

This paper presents a semi-supervised framework, based on Generative Adversarial Networks (GAN), for the classification of multi-spectral images. The semi-supervisd GAN network is trained on the weedNet dataset captured by an MVA from a sugar beet field. The performance of the system was evaluated using F1 score metric by varying the number of input channels and the amount of the labeled data. Results showed the F1 score of about 0.85 for two channels with 50% labeled data. Compared with weedNet paper [10] that utilized all the labeled data for training their decoder-encoder cascaded CNN for crop/weed classification, this paper demonstrated that even with limited labeled data the semi-supervised GAN network can classify the crop and weed with a relatively good accuracy. Future work includes testing the algorithm with other multispectral plant/weed datasets that contain more channels such as Infrared, Red edge, Red, Green and Blue to investigate the effect of each channel separately and in combination with each other on the classification performance.